\documentclass[twocolumn,showpacs,aps,superscriptaddress]{revtex4}
\usepackage{amsmath,amssymb,eucal,graphicx,bm}
\begin{document}
\title{Popularity-Driven Networking}
\author{E.~Ben-Naim} 
\affiliation{Theoretical Division and Center for Nonlinear Studies,
Los Alamos National Laboratory, Los Alamos, New Mexico 87545, USA}
\author{P.~L.~Krapivsky} 
\affiliation{Department of Physics, Boston University, Boston,
Massachusetts 02215, USA}
\begin{abstract}
We investigate the growth of connectivity in a network.  In our model,
starting with a set of disjoint nodes, links are added sequentially.
Each link connects two nodes, and the connection rate governing this
random process is proportional to the degrees of the two
nodes. Interestingly, this network exhibits two abrupt transitions,
both occurring at finite times.  The first is a percolation transition
in which a giant component, containing a finite fraction of all nodes,
is born.  The second is a condensation transition in which the entire
system condenses into a single, fully connected, component.  We derive
the size distribution of connected components as well as the degree
distribution, which is purely exponential throughout the
evolution. Furthermore, we present a criterion for the emergence of
sudden condensation for general homogeneous connection rates.
\end{abstract}
\pacs{89.75,Hc, 05.40.-a, 02.50.Ey, 05.20.Dd}
\maketitle

Networks are sets of nodes that are connected by links. Models for the
evolution of complex networks fall into two general classes: {\em
evolving graphs} where the number of nodes is fixed but the number of
links grows, and {\em growing networks} where the number of nodes and
the number of links both grow \cite{bb,jlr,dm,gc,mejn,ek,krb}. The
classic evolving random graph model, where pairs of randomly selected
nodes are repeatedly connected by links, captures the nucleation of a
giant connected component with a macroscopic number of nodes
\cite{pf,er}.  The preferential attachment model of network growth,
where newly added nodes connect to existing nodes with probability
that is proportional to the degree, yields the broad degree
distributions with power-law tails that characterize many real-world
complex networks \cite{has,ba}.

In this letter, we study random graphs that evolve according to a
preferential attachment mechanism. In our model, the number of nodes
is fixed but the number of connections grows. The ``popularity'' of
each node, as measured by the degree, governs the connection
process. This fusion between the two seminal network models of random
graphs and preferential attachment, is inspired by Facebook, the
immense online network of cyber-friends \cite{ek}.  In Facebook, new
connections are formed via friendship requests from one member of the
network to another; upon acceptance, the two become
friends. Preferential attachment implies that members seek and accept
friends according to popularity.

In our model for the popularity-driven growth of connectivity in a
network, the system consists initially of a set of disjoint nodes. We
consider the natural situation where the degree of a node, defined as
the number of existing connections, controls the rate by which two
nodes connect with each other. Specifically, a node with degree $i$
and a node with degree $j$ connect at a rate that is linear in $i$ and
linear in $j$ as well. We find that the degree distribution remains
purely exponential throughout the evolution. Thus, the preferential
attachment mechanism leads to opposite results in fixed and in growing
networks: the tail of the degree distribution is narrow in the former
but broad in the latter.

Our main result is that the system undergoes two finite-time
transitions. In the first transition, occurring at time $t_g$, a giant
connected component nucleates.  The giant component contains a finite
fraction of all nodes. In the second condensation transition, the
giant component takes over the entire system. In contrast with
classical random graphs, the entire network condenses into a single
component at a finite time $t_c$. We obtain analytically the degree
distribution and the size distribution of connected components when
the rate of connection between nodes with degree $i$ and nodes with
degree $j$ equals the product $(i+1)(j+1)$.  From these two
distributions, we obtain the finite transition times:
\begin{equation}
\label{tgtc}
t_g=1/3\qquad{\rm and}\qquad t_c=1.
\end{equation}
The rest of this letter includes two parts.  In the first, we study
how the degree of a node grows as a result of the networking process,
and in the second, we investigate how the size of a connected
component increases as a result of the very same process.

\begin{figure}[t]
\includegraphics[width=0.4\textwidth]{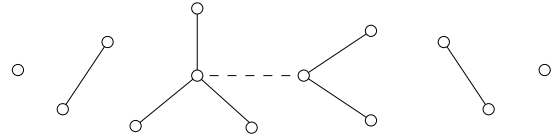}
\caption{Connection and subsequent aggregation.  The new link (dashed
line) connects two nodes with degrees $i=3$ and $j=2$. As a
consequence, two connected components with sizes $l=4$ and $m=3$
merge.}
\label{fig-link}
\end{figure}

In our evolving random graph model, the network consists of a set of
disjoint nodes at time $t=0$.  The network becomes connected through
the sequential addition of links: a node with degree $i$ and a node
with degree $j$ connect with rate $C_{i,j}$. As a result, the degree
undergoes the augmentation process (figure \ref{fig-link})
\begin{equation}
\label{connection}
(i,j)\stackrel{C_{i,j}}{\longrightarrow} (i+1,j+1).
\end{equation}
Since the two nodes are interchangeable, the connection rate is
symmetric, $C_{i,j}=C_{j,i}$.  We note that the classic random graph
corresponds to the uniform connection rate $C_{i,j}=1$ \cite{pf,er}.

Throughout this study, we implicitly take the infinite system size
limit. Let $n_j(t)$ be the degree distribution, that is, the fraction
of nodes with degree $j$ at time $t$. This quantity is normalized,
$\sum_j n_j=1$, and it obeys the rate equation
\begin{equation}
\label{nj-eq}
\frac{dn_j}{dt}=\nu_{j-1}n_{j-1} - \nu_j n_j.
\end{equation}
The initial condition is $n_j(0)=\delta_{j,0}$.  The quantity $\nu_j$
equals the total connection rate of nodes with degree $j$,
$\nu_j=\sum_i C_{i,j}\,n_i$.  

By summing the evolution equations \eqref{nj-eq}, we can verify that
the normalization is preserved, \hbox{$(d/dt)\sum_j
n_j=0$}. Similarly, by multiplying \eqref{nj-eq} with $j$ and summing
over all $j$, we find that the average degree, \hbox{$\langle j
\rangle =\sum_j jn_j$}, grows according to
\begin{equation}
\label{jav-eq-0}
\frac{d\langle j \rangle}{dt}=\sum_{i,j} C_{i,j}n_in_j.
\end{equation}
This equation reflects the connection process \eqref{connection}.  Since
every link connects two nodes, the average degree conveniently yields
the total density of links, $L$, via the simple relation $2L=\langle j
\rangle$.

In the preferential attachment model of network growth, the attachment
rate is linear in the degrees of the existing nodes. In this
``rich-get-richer'' mechanism, the connection probability is
proportional to the popularity. Hence, to model popularity-driven
networking in evolving graphs, we restrict our attention to rates that
are linear in both $i$ and $j$, $C_{i,j}=(i+a)(j+a)$. Often used in
preferential attachment, the offset $a$ allows us to start with a set
of disjoint nodes, the natural initial condition.  Here, we
investigate the rate
\begin{equation}
\label{rate-connection}
C_{i,j}=(i+1)(j+1).
\end{equation}
We verified that the qualitative behavior, including the the sudden
condensation transition, extends to all $a$ \cite{dimer}.

By substituting the rate \eqref{rate-connection} into equation
\eqref{jav-eq-0}, we see that the average degree obeys the closed
equation
\begin{equation}
\label{jav-eq}
\frac{d\langle j \rangle}{dt} = \big(\langle j \rangle+1\big)^2. 
\end{equation}
We obtain the average degree by solving this equation, subject to the
initial condition $\langle j\rangle\big|_{t=0} =0$,
\begin{equation}
\label{jav-sol}
\langle j \rangle = \frac{t}{1-t},
\end{equation}
when $t<1$.  Hence, the average degree diverges in finite time.

From the average degree \eqref{jav-sol}, we find that the connection
rate $\nu_j$ is linear in the degree, \hbox{$\nu_j=(j+1)(1-t)^{-1}$}.
Therefore, the degree distribution obeys the linear evolution equation
\begin{equation}
\label{nj-eq-1}
(1-t)\frac{dn_j}{dt}=j\,n_{j-1} - (j+1)\, n_j.
\end{equation}
We solve these equations recursively, starting with the initial
condition $n_j(0)=\delta_{j,0}$, and obtain \hbox{$n_0=1-t$},
\hbox{$n_1=(1-t)t$}, \hbox{$n_2=(1-t)t^2$}, etc. Generally one finds 
that the degree distribution is purely exponential,
\begin{equation}
\label{nj-sol}
n_j=(1-t)\,t^j.
\end{equation} 
Surprisingly, the degree distribution vanishes in finite time:
$n_j(t_c)=0$ for all $j$ with $t_c=1$. Consequently, the degree of a
node is finite if and only if $t<t_c$.  This behavior is a signature
of the continuous condensation transition that occurs at the
condensation time $t_c$. As shown below, the entire network condenses
into a single, fully connected, component at this time.

The difference between growing and fixed networks is remarkable.
Networks that grow by preferential attachment have broad degree
distributions with power-law tails \cite{has,ba,krl,dms,bk09}, but in
a fixed network, preferential attachment leads to a narrow degree
distribution with exponential tail. On its own, the rich-get-richer
mechanism does not generate broad tails, but rather, it is the
combination of a growing network and preferential attachment that
leads to a broad distribution of degrees. We note that exponential
degree distributions occur in a variety of complex networks including
power, transportation, and social communication networks
\cite{lj,dlcw}.

Our evolving graph consists of multiple connected components
(``clusters'') which undergo binary aggregation as a result of the
connection process \eqref{connection}.  Symbolically, such an
aggregation process \cite{fl,aal} can be represented as (figure
\ref{fig-link})
\begin{equation}
\label{aggregation}
[l] + [m]\stackrel{K_{l,m}}{\longrightarrow} [l+m],
\end{equation}
where $l$ and $m$ are the number of nodes in the two merging clusters.
The rate of aggregation between a {\em finite} cluster of size $l$ and
a {\em finite} cluster of size $m$ is
\begin{equation}
\label{rate-aggregation}
K_{l,m}=(3l-2)(3m-2).
\end{equation}
To obtain this rate, we note that a cluster with $k$ nodes has $k-1$
links. (This relation is valid for clusters with tree structure, and
indeed, nearly all finite clusters are trees in the infinite size
limit \cite{krb}.)  Equation \eqref{rate-aggregation} follows from the
connection rate \eqref{rate-connection}, together with the fact that
the sum of the degrees in the cluster equals twice the number of
links.

The total density of clusters, $c$, follows from the total density
of links, $L$, when all clusters are finite.  Since every link reduces
the number of clusters by one, we have $c(t)=1-L(t)$. Using
$2L=\langle j\rangle$, we conclude $c=1-\langle j\rangle/2$, and from
\eqref{jav-sol}, we obtain the cluster density
\begin{equation}
\label{ct-sol}
c(t)=\frac{2-3t}{2(1-t)}\,.
\end{equation}
As shown below, this relation holds when $t<t_g$ with $t_g$ given in
\eqref{tgtc}.

The density $c_k(t)$ of clusters with size $k$ at time $t$ obeys the
master equation
\begin{eqnarray}
\label{ck-eq}
\frac{dc_k}{dt} \!= \!\tfrac{1}{2}\!\!\!
\sum_{l+m=k}\!\!\!(3l\!-\!2)(3m\!-\!2)c_l c_m 
- \langle j\!+\!1 \rangle(3k\!-\!2)c_k,
\end{eqnarray}
where $\langle j +1 \rangle =(1-t)^{-1}$. The initial condition is
\hbox{$c_k(0)=\delta_{k,1}$}.  The gain term is {\em nonlinear}, and
it directly reflects the aggregation process \eqref{aggregation} with
the product rate \eqref{rate-aggregation}. The loss term, on the other
hand, is {\em linear} in the cluster-size density. The loss rate
equals the product between the sum of all degrees (shifted by one) in
the {\em cluster}, and the average degree (shifted by one) of all
nodes in the entire {\em system}. In this form, the master equation is
valid at all times, whether all clusters are finite or whether
macroscopic clusters exist as well \cite{bk11}.

Let $M_n=\sum_n k^n c_k$ be the $n$th order moment of the size
distribution. The zeroth moment gives the total cluster density,
$M_0\equiv c$, and the first moment, $M_1$, yields the total mass of
finite components. Furthermore, the second moment satisfies
\begin{eqnarray*}
\frac{dM_2}{dt} &=& (3M_2-2M_1)^2\\
&+&(3M_3-2M_2)\left[(3M_1-2M_0)-\langle j+1\rangle\right],
\end{eqnarray*}
subject to the initial condition $M_2(0)=1$. We obtain this equation
by multiplying \eqref{ck-eq} by $k^2$ and summing over all $k$. Let's
assume that finite clusters contain all the mass, $M_1=1$. In this
case, the cluster density \eqref{ct-sol} gives
\hbox{$3M_1-2M_0=(1-t)^{-1}$}, and as a consequence, the second
moment satisfies the closed equation \hbox{$dM_2/dt =
(3M_2-2)^2$}. Therefore,
\begin{equation}
M_2=\frac{1-2t}{1-3t}\,,
\end{equation}
for $t<1/3$.  The divergence of the second moment shows that a
percolation transition \cite{sa} occurs at time \hbox{$t_g=1/3$} as
stated in \eqref{tgtc}.  The critical density of links
\hbox{$L_g=L(t_g)=1/4$} is smaller than the value $L_g=1/2$
corresponding to the classic random graph \cite{er}.  

When $t<t_g$, the system is in a non-percolating phase and finite
clusters contain all the mass, $M_1=1$. Otherwise, the system is in a
percolating phase, where a macroscopic cluster, the giant component
\cite{jklp}, contains a finite fraction of all nodes and $M_1<1$.

To find the mass of the giant component, we analyze the cluster-size
distribution.  First, we solve \eqref{ck-eq} recursively for small
clusters. We can confirm that the density of minimal-size clusters
(``monomers'') equals the fraction of isolated nodes,
\hbox{$c_1=n_0=1-t$}.  The densities of ``dimers'' and ``trimers'' are
\hbox{$c_2 = \tfrac{1}{2}t(1-t)^3$} and \hbox{$c_3 = t^2(1-t)^5$}.  We
therefore expect the general form
\begin{equation}
\label{ck-sol}
c_k = A_k\, t^{k-1} (1-t)^{2k-1}\,.
\end{equation}
By substituting this expression into \eqref{ck-eq}, we obtain a
recursion equation for the coefficients $A_k$
\begin{equation}
\label{Ak-eq}
2(k-1)A_k = \sum_{l + m=k}(3l-2)(3m-2) A_l A_m, 
\end{equation}
for $k>1$, subject to $A_1=1$. We now make the transformation
$B_k=(3k-2)A_k$. The coefficients $B_k$ satisfy a second recursion
relation
\begin{equation*}
2(k-1)B_k = (3k-2)\sum_{l + m=k}B_l B_m,
\end{equation*}
for $k>1$, subject to $B_1=1$.  From this recursion, we conclude that
the generating function \hbox{$B(x)=\sum_k B_k\, x^k$}, obeys the
differential equation \hbox{$x\frac{dB}{dx}=\frac{B(1-B)}{1-3B}$}. We
now integrate this differential equation subject to the constraint
$\lim_{x\to 0}x^{-1}B=1$, and find that $B(x)$ satisfies the cubic 
equation
\begin{equation}
\label{bx-eq}
B(1-B)^2=x.
\end{equation}

We obtain the coefficients $B_k$ by using the Lagrange inversion
method,
\begin{eqnarray*}
B_k&=&\frac{1}{2\pi i}\oint \frac{B}{x^{k+1}}dx \\
   &=&\frac{1}{2\pi i}\oint \frac{B(1-B)(1-3B)}{B^{k+1}(1-B)^{2(k+1)}}dB \\
   &=&\frac{1}{2\pi i}\oint \frac{1-3B}{B^k(1-B)^{2k+1}} dB\\
   &=&\frac{1}{2\pi i}\oint \left[\sum_{n=0}^\infty \binom{2k+n}{n}\left(B^{n-k} - 3B^{n+1-k}\right)\right]dB\\
   &=&\binom{3k-1}{k-1}-3\binom{3k-2}{k-2}=\frac{(3k-2)!}{k!(2k-1)!}.
\end{eqnarray*}
In the second line, we used equation \eqref{bx-eq}, and replaced the
integration over $x$ with integration over $B$, by using
\hbox{$dx=(1-B)(1-3B)dB$}. Finally, we utilize the identity
$(1-B)^{-m}=\sum_{n=0}^\infty \binom{m+n-1}{n}B^n$, to determine the
coefficients $A_k$,
\begin{equation}
\label{Ak-sol}
A_k=\frac{(3k-3)!}{k!(2k-1)!}.
\end{equation}
The integer sequence $2A_k=\{2,1,2,6,22,91,\ldots\}$ also arises in 
 planar maps \cite{njas,wtt}.

The tail of the cluster-size distribution is a product of a power law
and an exponential,
\begin{equation}
\label{ck-tail}
c_k \simeq \frac{1}{\sqrt{12 \pi}}\, k^{-5/2}\,e^{-k/k_*}, 
\end{equation}
for $k\to\infty$, with $k_*^{-1}=\ln
\frac{t_g(1-t_g)^2}{t(1-t)^2}$. This tail follows from the Stirling
formula and the coefficients \eqref{Ak-sol}.  Hence, at the
percolation transition, the distribution is purely algebraic,
$c_k(t_g)\sim k^{-5/2}$ \cite{zhe}. Otherwise, the tail is power law
at small scales, $k\ll k_*$, but exponential at large scales, $k\gg
k_*$. The characteristic scale diverges, $k_*\simeq \tfrac{4}{27}
(t-t_g)^{-2}$, as $t\to t_g$.

\begin{figure}[t]
\includegraphics[width=0.4\textwidth]{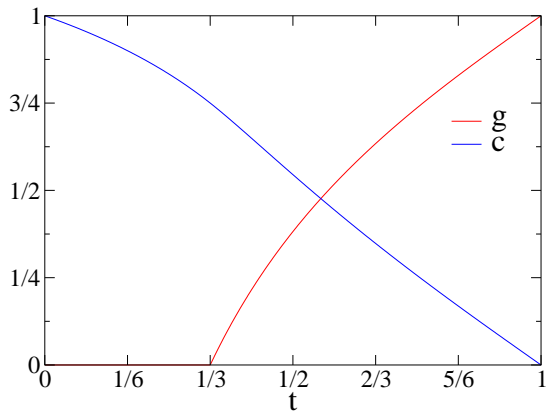}
\caption{The mass of the giant component, $g$, and the total density
of connected components, $c$, versus time $t$.}
\label{fig-gt}
\end{figure}

Note that the time dependence of the cluster-size density
\eqref{ck-sol} enters primarily through the quantity \hbox{$t(1-t)^2$}
as $c_k \propto [t(1-t)^2]^k$.  This observation tells us that for an
auxiliary time $\tau$, that is connected to the physical time $t$ via
the cubic equation
\begin{equation}
\label{tau}
\tau(1-\tau)^2 = t(1-t)^2, 
\end{equation}
the densities $c_k(t)$ and $c_k(\tau)$ are related by the duality
relation
\begin{equation}
\label{dual}
c_k(t)=c_k(\tau)\left(\frac{1-t}{1-\tau}\right).
\end{equation}
When $t<t_g$, equation \eqref{tau} has only the trivial solution
$\tau=t$, while for $t_g<t<t_c$ there is additionally a second
nontrivial root $\tau<t_g$.  Consequently, for all $t_g<t<t_c$ we can
choose the nontrivial root $\tau<t_g$ of \eqref{tau} and then, the
duality relation \eqref{dual} conveniently specifies the cluster-size
distribution in the percolating phase in terms of its counterpart in
the non-percolating phase.  By multiplying \eqref{dual} by $k$ and
summing over all $k$ we see that
\hbox{$M_1(t)/(1-t)=M_1(\tau)/(1-\tau)$}, and since $M_1(\tau)=1$, we
immediately get the nontrivial first moment in the percolating phase,
$M_1=(1-t)/(1-\tau)$. In summary, the mass of finite clusters is
\begin{equation}
\label{M1-sol}
M_1(t) = 
\begin{cases}
1                                                       & 0\leq t\leq t_g,\\
\frac{1-t}{1-\tau}                                & t_g\leq t\leq t_c,\\
0                                                        & t\geq t_c.
\end{cases}
\end{equation}
The mass of the giant component, $g$, equals the complementary mass,
$g=1-M_1$ (see Fig.~\ref{fig-gt}). Post-percolation, this quantity
grows linearly, $g(t)\simeq 3(t-t_g)$, as $t\downarrow t_g$, and
furthermore, $1-g(t)\simeq t_c-t$ as $t\uparrow t_c$. At time $t_c=1$,
the giant component takes over the entire graph, and hence, the system
condenses into a single component. Remarkably, the cluster-size
density and the fraction of nodes with finite degrees vanish {\em
simultaneously}, $c_k(t_c)=n_j(t_c)=0$.

For completeness, we mention that the generating function, ${\cal
C}(z,t)=\sum_k c_k(t)\,z^k$, can be written as an explicit function of
time using the hypergeometric function,
\begin{equation*}
{\cal C}(z,t)=\frac{1}{3t(1-t)}
\left[F\left(-\tfrac{2}{3},-\tfrac{1}{3};
\tfrac{1}{2};\tfrac{3^3}{2^2}t(1-t)^2\,z\right)-1\right].
\end{equation*} 
Moments of the distribution can also be written explicitly \cite{m2}.
The density $c(t)\equiv {\cal C}(z=1,t)$ is plotted in figure
\ref{fig-gt}, and one can verify that this quantity matches
\eqref{ct-sol} in the non-percolating phase.

An unusual feature of our network connection process is that both the
percolation transition and the condensation transition occur at finite
times. Let us consider the uniform connection rate $C_{i,j}=9$,
corresponding to the classic random graph. For this evolving graph,
the percolation time is finite, $t_g=1/9$, but the condensation time
is divergent, $t_c=\infty$ (in a finite system, the condensation time
is logarithmic in the total number of nodes)
\cite{krb}. Interestingly, the corresponding aggregation rate,
$K_{l,m}=(3l)(3m)$, is {\em larger} than the aggregation rate
\eqref{rate-aggregation}, yet the latter, smaller, rate produces
faster condensation!  Of course, the aggregation rate $K_{l,m}$
corresponds only to {\em finite} clusters, and it does not apply to
the giant component.  The rich-get-richer mechanism accelerates the
rate by which the giant component engulfs finite components, and this
feature is ultimately responsible for the finite-time condensation.

We also considered a general class of evolving graphs with homogeneous
connection rates, and we now briefly discuss the purely algebraic
rates
\begin{equation}
C_{ij}=(ij)^\alpha\,.
\end{equation}
The case $\alpha=0$ corresponds to ordinary random graphs, and the
case $\alpha=1$, to popularity-driven networking. From
\eqref{jav-eq-0}, the average degree grows according to
\begin{equation}
\label{jav-eq-gen}
\frac{d\langle j\rangle}{dt}\sim \langle j^\alpha\rangle^2\sim \langle
j\rangle^{2\alpha},
\end{equation}
where we assumed the scaling behavior $\langle j^\alpha\rangle\sim
\langle j\rangle^\alpha$.  From this rate equation, we find the
scaling behaviors
\begin{equation}
\langle j\rangle\sim
\begin{cases}
t^{1/(1-2\alpha)} & \alpha<1/2,\\
e^{{\rm const.}\times t}                    & \alpha=1/2,\\
(t_c-t)^{-1/(2\alpha-1)}  &1/2<\alpha\leq 1.
\end{cases}
\end{equation}
When $\alpha<1/2$, the average degree grows algebraically with
time. Furthermore, the percolation time is finite, but the
condensation time is infinite. In the marginal case $\alpha=1/2$,
the degree grows exponentially with time. When $1/2<\alpha\leq 1$, the
percolation time is finite, and the condensation time is finite as
well. The average degree diverges as the condensation transition is
approached.

Finally, when $\alpha>1$, the scaling assumption used in
\eqref{jav-eq-gen} is no longer valid, and condensation becomes
instantaneous: $t_c=t_g=0$ \cite{bk12}.  This phenomenon is analogous
to the instantaneous gelation that occurs in aggregation
\cite{pgjv,bcsz} and exchange-driven growth \cite{bk03}.

In conclusion, we studied the growth of connectivity in an evolving
graph. In our model, the degree controls the connection process as the
probability that a node forms a new connection is proportional to the
total number of its existing connections. We find that the system
undergoes two continuous transitions, both occurring at finite
times. In the first percolation transition, a macroscopic connected
component nucleates. In the second condensation transition, the
network becomes fully connected. We obtained analytically the degree
distribution and the size distribution of connected components.  We
find that in a fixed network, the degree distribution is exponential,
and therefore, the average degree fully characterizes the entire
distribution.

Our theoretical analysis relies heavily on the fact that the
connection rate is linear in the degree.  For such connection rates,
the rate by which components merge is a bilinear function of size.
Consequently, the size distribution of connected components is
analytically tractable.  Linear connection rates also have
computational advantages.  Popularity-driven networking can be studied
using a convenient {\em random link algorithm}. In this numerical
simulation, we choose two links at random. For each link we pick one
of the two nodes connected to it, and then, connect these two
nodes. With this implementation, we select nodes with probability that
is proportional to the degree \cite{krb}, and this procedure can be
generalized to arbitrary linear rates.  The random link algorithm has
optimal efficiency as the computational cost is linear in the number
of links, and it can be used to study additional properties of our
evolving graph. Among a host of possibilities, we mention structural
properties of connected components such as their cycle structure
\cite{jklp}, and extremal properties such as statistics of the largest
connected component.

\smallskip
This research was supported by DOE grant DE-AC52-06NA25396 and NSF
grant CCF-0829541.


\begin{thebibliography}{99}
\bibitem{dimer} The product rate $C_{i,j}=ij$ leads to qualitatively
       similar results if initially there are no isolated nodes.
\bibitem{m2} 
      The fraction of nodes in finite components is given by
      $M_1(t)=\frac{2}{\sqrt{3t}}\sin\left[\frac{1}{3}\sin^{-1}\left(\frac{3\sqrt{3t}\,(1-t)}{2}\right)\right]$.
\end{thebibliography}
\end{document}